\documentclass[preprint]{revtex4-1}
\usepackage{amsmath}
\usepackage{amssymb}
\usepackage{graphicx}
\begin{document}

\title{Fabrication of an InGaAs spin filter by implantation of paramagnetic centers} 

\author{C. T. Nguyen}
\author{A. Balocchi}
\author{D. Lagarde}
\author{T. T. Zhang}
\author{H. Carr\`ere}
\author{S. Mazzucato}
\author{P. Barate}
\affiliation{Universit\'{e} de Toulouse, INSA-CNRS-UPS, LPCNO, 135 avenue de Rangueil, 31077 Toulouse, France}
\author{E. Galopin}
\author{J. Gierak}
\author{E Bourhis}
\author{J. C. Harmand}
\affiliation{CNRS-LPN, Route de Nozay, 91460 Marcoussis, France}
\author{T. Amand}
\author{X. Marie}
\affiliation{Universit\'{e} de Toulouse, INSA-CNRS-UPS, LPCNO, 135 avenue de Rangueil, 31077 Toulouse, France}

\begin{abstract}
We report on the selective creation of spin filtering regions in non-magnetic InGaAs layers by implantation of Ga  ions by Focused Ion Beam. 
We demonstrate by photoluminescence spectroscopy that spin dependent recombination (SDR) ratios as high as 240\% can be achieved in the implanted areas.
The optimum implantation conditions for the most efficient SDR is determined by the systematic analysis of  different ion doses spanning four orders of magnitude.
The application of a weak external magnetic field leads to a sizable enhancement of the SDR ratio from the spin polarization of the nuclei 
surrounding the polarized implanted paramagnetic defects.
\end{abstract}
\pacs{85.75.-d 61.72.uj 78.47.D-}
\maketitle 
The defect engineered approach to the spin filtering problem in dilute nitrides has been  drawing much attention 
in the spintronics community thanks to their remarkable spin filtering efficiency even at room temperature. By relying on the dependence of the
recombination time on the relative spin orientation
of photogenerated carriers on paramagnetic  centers (the spin dependent recombination - SDR), the photo-created conduction band electron spin polarization can be regenerated or amplified in an few tens of picoseconds (figure 1b).
Althought the SDR has been already known in semiconductors for over 30 years in silicon~\cite{lepine_1970,lepine_1972,wosinski_1977,solomon_1976} and (Al)GaAs~\cite{weisbuch_1974,paget_1984} it manifests itself with extreme effectiveness in dilute nitrides.
\begin{figure}[rh!]
\includegraphics[width=0.5\textwidth]{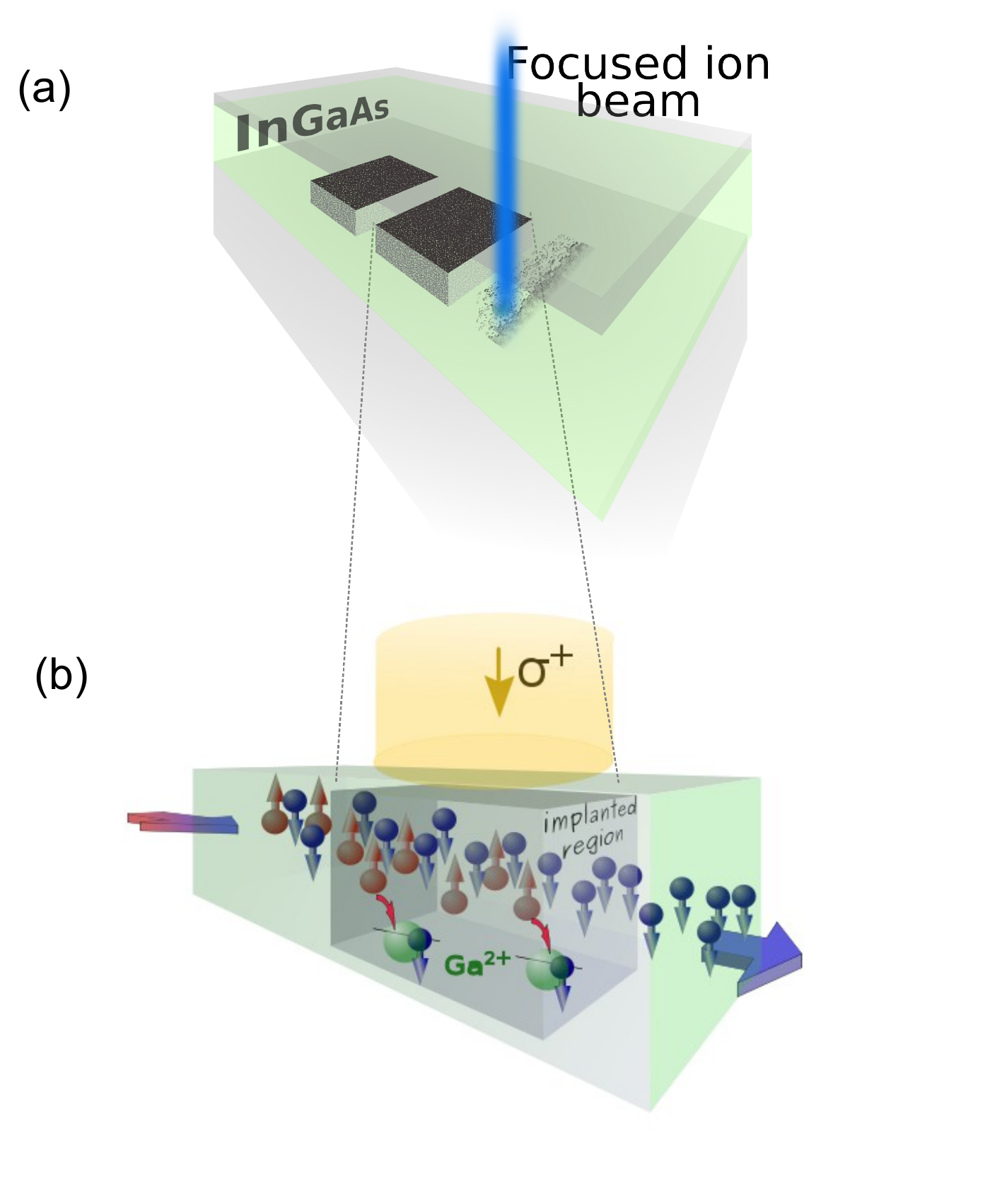}
\caption{(a) Schematic picture of the fabrication of the spin filtering zones by focused ion beam implantation of Ga ions. Seven different 200 $\mu$m side square regions were implanted in the 50 nm thick InGaAs layer with surface doses ranging from  1.8$\times$10$^8$ cm$^{-2}$  to 1.8$\times$10$^{12}$ cm$^{-2}$. (b) Representation of the operation of the device. The spin filtering region is activated by circularly polarized light ($\sigma^+$) which favors the spin polarization of electrons resident on the paramagnetic centers. A flow of spin-unpolarized CB electrons traversing the region will experience the amplification of its spin polarization degree due to the selectivity of Ga$^{2+}$ defect capture properties of oppositely spin oriented  CB electrons compared to the residents ones. The outwards CB electrons flow is almost fully spin polarized.}
\label{figure_1}
\end{figure}
\begin{figure}
\includegraphics[width=0.5\textwidth]{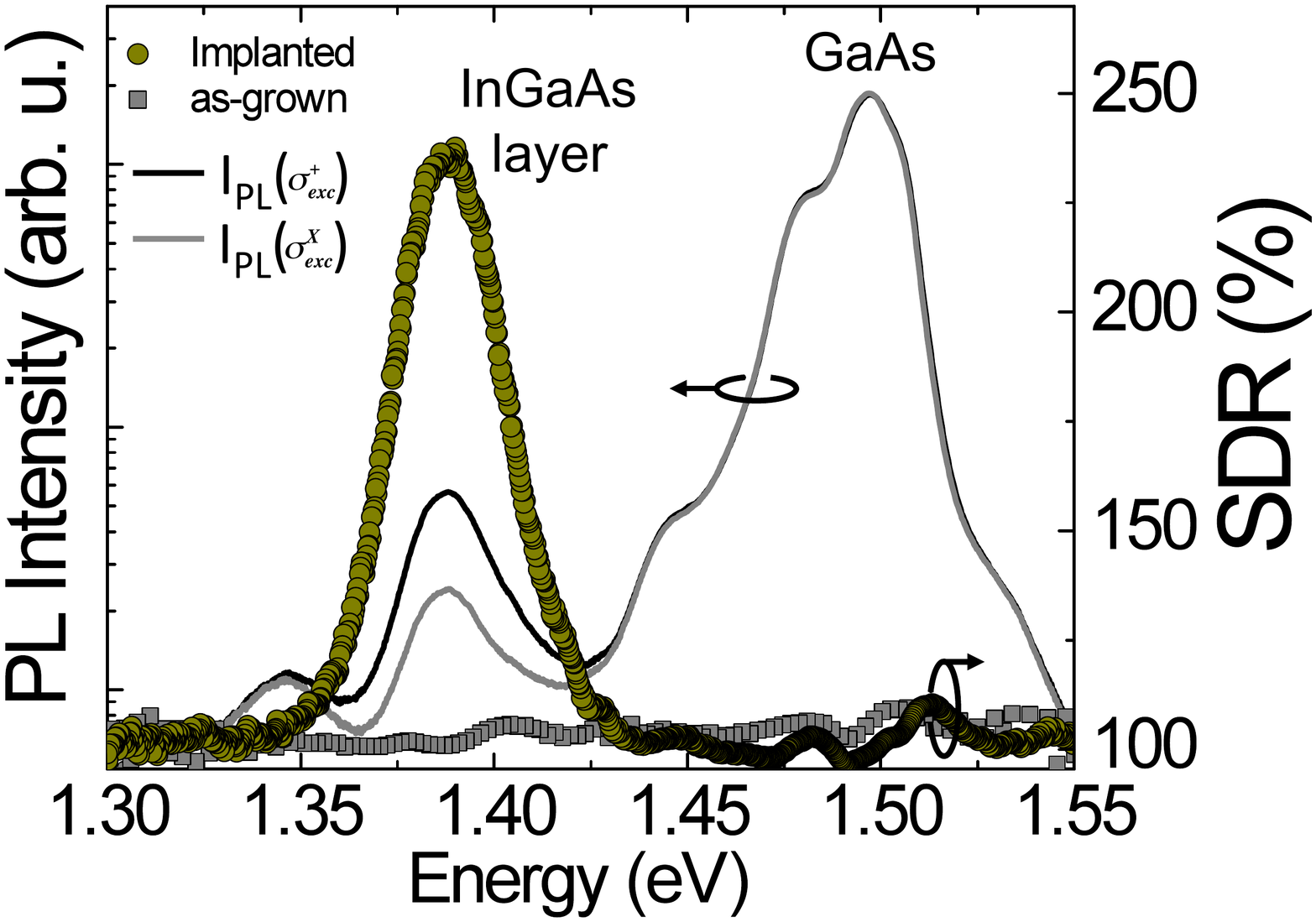}
\caption{Photoluminescence intensity recorded under a circular (black line) and linear (gray line) excitation in the implanted region ($P_{exc}$ = 10 mW, $T$= 25 K). The  circles reproduce the corresponding SDR ratio. The  squares report the SDR ration measured outside the implanted region under the same conditions.}
\label{figure_2}
\end{figure}
\begin{figure}
\includegraphics[width=0.5\textwidth]{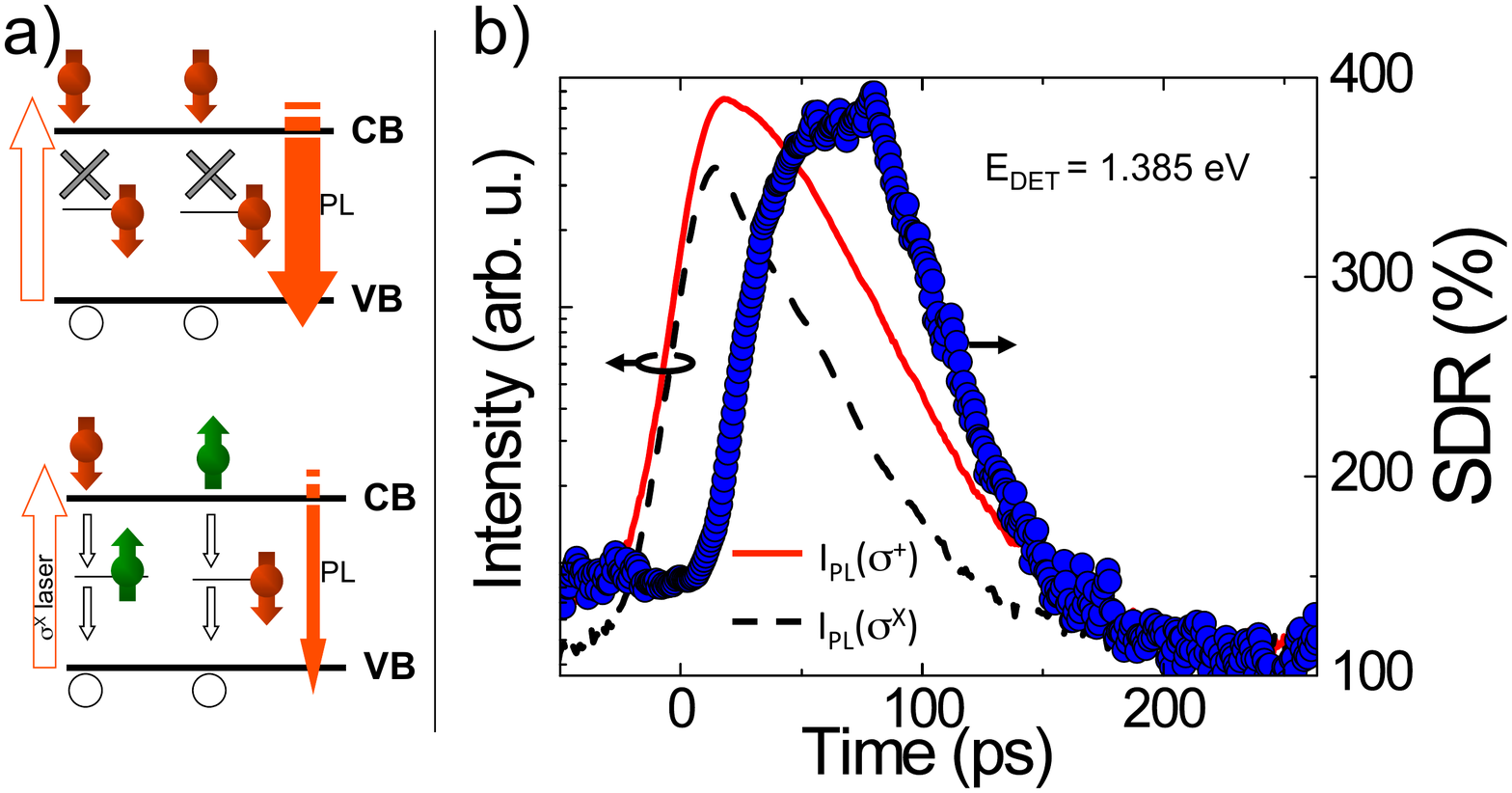}
\caption{(a) Conduction band capture and recombination mechanisms at the origin of the SDR effect under an optical excitation of different polarization. The downwards arrows represent the possible recombination processes for the conduction band electrons either by band to band recombination (filled PL arrow) or through capture by a deep paramagnetic center. (b) The photoluminescence dynamics measured at the emission peak (E$_{det}$= 1.385 eV) of the implanted InGaAs layer and the corresponding dynamics of the SDR ratio.}
\label{figure_3}
\end{figure}
Subsequently to the demonstrations in (In)GaAsN epilayers and quantum wells\cite{kalevich:455,lombez:252115,kalevich:174,lagarde:208} by  photoluminescence spectroscopy, the SDR in dilute nitrides has been also demonstrated  by photoconductivity~\cite{zhao:241104}.
More recently the enhancement of the SDR effect under a weak longitudinal magnetic field (Faraday geometry) has been evidenced~\cite{Kalevich_B_field_1,Kalevich_B_field_2}, giving signatures of nuclear polarization effects at room temperature. Moreover, optically detected magnetic resonance experiments have demonstrated that it is nitrogen-induced gallium self-interstitial Ga$^{2+}$
that plays the role of the deep paramagnetic center~\cite{wang_nature_materials}.\\
The fabrication of SDR active regions is usually achieved by epitaxial growth. SDR active quantum wells or epilayers can thus be produced with in-plane uniformity but without easy engineering possibilities of specific active region patterns.
In addition, the introduction of even small quantities of nitrogen in (In,Al)GaAs layers leads to two other significant modifications. On one side a giant reduction of the band gap energy ($\approx$ 150 meV  per percent of N)~\cite{weyers_red_1992,wu_band_2002} making it impossible to dissociate the SDR efficiency from the PL emission wavelength~\cite{Zhao_JOP}. On the other side, tensile  strain is introduced due to the modification of the lattice constant compared to GaAs.\\
The aim of this work is to put forward a proof of concept demonstrating the fabrication of an InGaAs spin filter without any nitrogen introduction.\\
We report in this Letter on a method allowing for the creation of spin filtering active regions with nanometer size in-plane resolution in (In)GaAs layers.  By implanting with Ga ions squared regions of nitrogen-free semiconductor layers by focus ion beam,  we demonstrate  the selective creation of spin filtering regions exhibiting SDR ratio values as high as 240\%. This method  avoids the giant band-bowing characteristic of the dilute nitrides~\cite{weyers_red_1992} and allows the creation of SDR active regions with arbitrary patterns and efficiency by controlling the implantation conditions. As it relies on the post-growth creation of  interstitial paramagnetic centers, this technique is in principle transferable to other (In,Al)GaAs compounds and allows for the independent tuning of the emission wavelength and SDR efficiency.\\
The sample under study was grown by molecular beam epitaxy on a semi-insulating (001)-oriented GaAs substrate.
The 50 nm thick In$_{0.09}$Ga$_{0.91}$As active layer was grown on top of  100 nm Al$_{0.25}$Ga$_{0.75}$As barrier and 10 nm GaAs spacing layers. The sample
 was then covered by a 5 nm thick GaAs cap layer (figure 1a). Seven different 200 $\mu$m side square regions were uniformly implanted, using a 35 keV focused ion beam probe
  transporting  3 pA. The digital scanning speed, dwell times and pixel-to-pixel distance were adjusted to provide surface doses ranging from 
  1.8$\times$10$^8$ cm$^{-2}$  to 1.8$\times$10$^{12}$ cm$^{-2}$. The acceleration energy of the Ga ions and the InGaAs layer distance form the surface were chosen in order to selectively create the deep paramagnetic
centers only in the InGaAs part while maintaining good optical qualities.\\
  Time-resolved spectroscopy was employed to evidence the SDR phenomenon at low temperature.
A mode-locked Ti:sapphire laser with 1.5 ps pulse width and 80MHz repetition frequency
was used to excite the sample. The excitation beam, propagating parallel to the growth axis,  was focused to a 30 $\mu$m spot diameter and the excitation energy was set to 1.57 eV. This energy was chosen to induce absorption in both the GaAs spacing layer and in the implanted InGaAs regions in order to evidence the selectivity of the ion implantation. An average power ranging from 1 to 30 mW was used.
The laser light was either circularly ($\sigma^+$) or linearly ($\sigma^X$) polarized and the total photoluminescence (PL) collected in a backscattering
 geometry, dispersed  by an imaging spectrometer and its dynamics measured by a streak camera with an overall time resolution of 8 ps. 
The SDR ratio is defined here as SDR=$I^+/I^X$ where $I^+$ and $I^X$ denote respectively the total PL intensity detected under circular or linear excitation.\\
We present here the result obtained on the sample implanted with 4.5$\times$10$^9$ cm$^{-2}$ ion density at T=25 K in order to maintain a strong PL intensity. Qualitatively similar results were obtained on all the other implanted regions except for the two highest  doses for which the ion-induced damages were too important to observe a clear luminescence.
Figure 2 reports the time integrated PL intensity under circular and linear excitation for the implanted region. The PL spectrum contains features related  both to the GaAs and InGaAs layers.
The main peak centered on 1.50 eV is identified as resulting from the near band edge and defect-related emission from GaAs. The spectral feature centered at 1.385 eV is unambiguously assigned to the luminescence emitted by the InGaAs layer.
Figure 2 reproduces additionally the corresponding SDR ratio (circle). A much  stronger PL intensity (SDR=240\%) is recorded for the implanted InGaAs layer 
under a circularly polarized excitation with respect to a linearly polarized one. The PL enhancement under circularly polarized excitation is absent (SDR=100\% within the experimental ucertainties) from the GaAs  and un-implanted InGaAs (squares) related emissions. The measured SDR value is consistent to what was reported for dilute nitride (In)GaAsN systems~\cite{lagarde:208,kunold_giant_2011}.\\ 
The principle of the SDR process mediated by a  paramagnetic  spin-filtering defect is schematically presented in figure 3a.
Under a circular excitation, spin polarized electrons are photogenerated in the conduction band (CB). The fast capture of these electrons initiates the dynamical spin polarization of the electrons resident on the paramagnetic defects which favors the same spin orientation as the CB electrons. The Pauli exclusion principle then prevents  any subsequent 
capture of CB electrons on the paramagnetic centers producing the band-to-band PL intensity enhancement. In this regime, a CB electron subject to a spin flip is immediately captured by the spin polarized deep centers. A strong spin polarized CB electron population can thus be sustained through this spin filtering mechanism.\\
Under linearly polarized excitation, no CB spin polarization can be created thus maintaining the original random  orientation of the paramagnetic centers' spins. The spin insensitive fast CB electrons capture on the unpolarized defect can persist giving rise to a weaker band-to-band PL.
The dynamics of the SDR effect is evidenced in figure 3b where the SDR ratio temporal evolution is reported (circles) as measured at the peak of the PL intensity related to the InGaAs layer.
\begin{figure}
\includegraphics[width=0.5\textwidth]{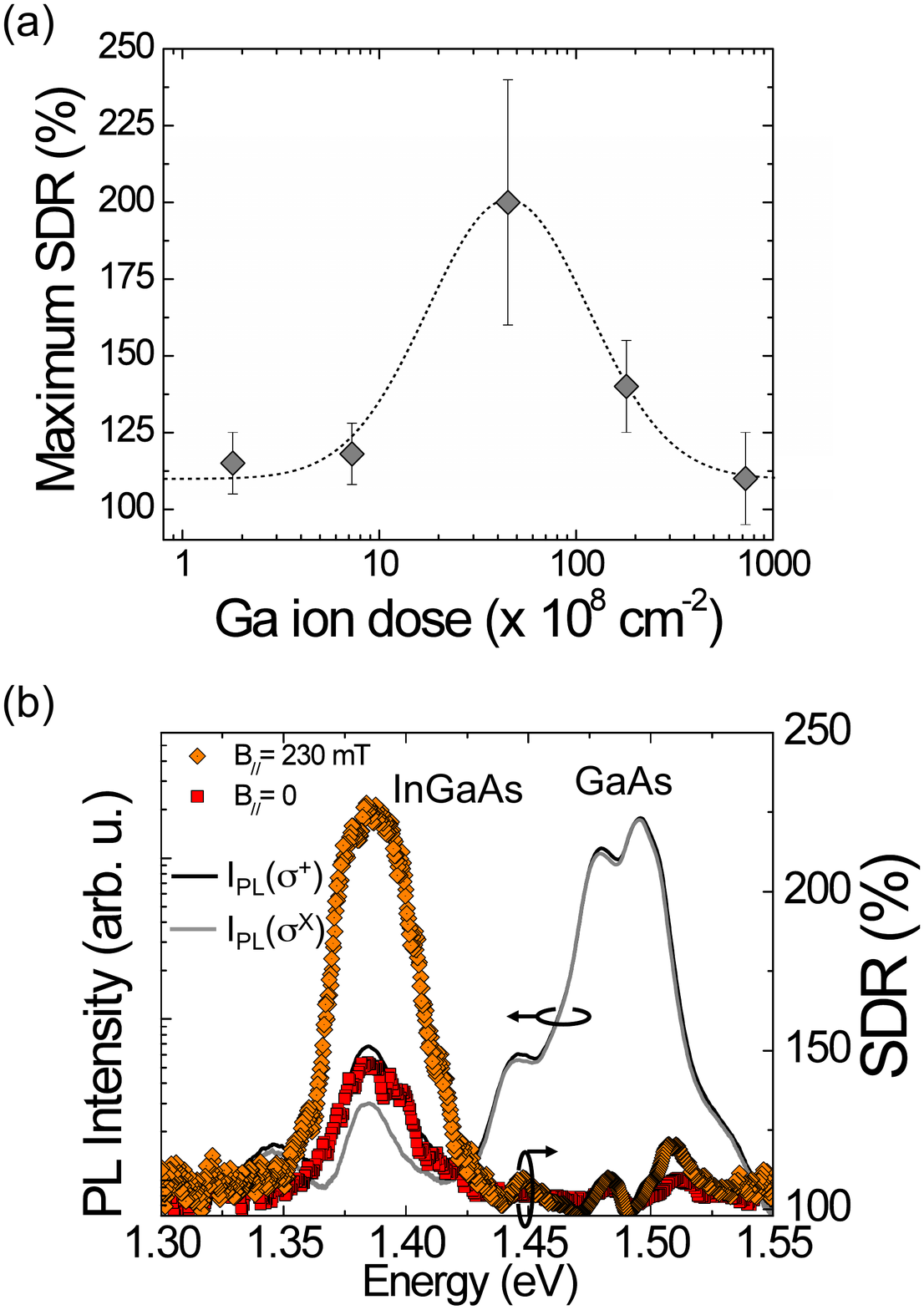}
\caption{(a) The maximum SDR ratio measured as a function of the ion implantation dose. The dotted line is a guide to the eyes. (b) The photoluminescence intensity recorded under a circular (black line) and linear (gray line) excitation in the implanted region ($P_{exc}$ = 5 mW, $T$= 25 K) under an external magnetic field $B_{\\parallel}$= 230 mT in Faraday geometry. The triangles reproduce the corresponding SDR ratio. The circles present the SDR ratios measured under the same conditions without the external magnetic field.}
\label{figure_4}
\end{figure}
A rapid  increase of the SDR ratio is observed reaching values as high as 380\% as the spin filtering regime is progressively reached in the first picoseconds after the excitation. The subsequent SDR decline reveals the onset of an insufficient CB population and defect spin  polarization necessary to sustain the process due to the band-to-band recombination~\cite{Zhao_JOP}.\\
All these observations are consistent with the creation of paramagnetic centers by Ga ion implantation.\\

Figure 4a  reports the spin dependent recombination ratio measured at the excitation intensity producing the maximum SDR value~\cite{Zhao_JOP} as a function of the implantation dose. For each dose, the errors bars are deduced from the inhomogeneity of the
 measured SDR ratio as observed in the same implanted square regions.
The Ga ion dose which produces the highest SDR ratio is 5$\times$ 10$^{9}$ cm$^{-2}$. At higher doses, the competition between the paramagnetic centers and  other ion-induced defects progressively reduces the SDR effect~\cite{kunold_giant_2011}. This is confirmed by the important degradation of the PL intensity for doses above 10$^{10}$ cm$^{-2}$. Indeed a complete disappearance of
any measurable PL signal is obtained at the two highest implantation doses used in this work.\\
Taking into consideration the optimum density of Ga ions (5$\times$10$^{9}$ cm$^{-2}$) determined here and assuming for simplicity 100\% efficiency for the creation of Ga$^{2+}$ interstitial by focused ion beam in the 50 nm thick InGaAs layer, we can estimate that the  density of paramagnetic centers created is in the order of  10$^{15}$ cm$^{-3}$ which is in good agreement with previous works on dilute nitrides~\cite{lagarde:208}. This suggests the assignment of the nature of the paramagnetic centers to Ga$^{2+}$ interstitials.\\
This statement is further supported by the measurement of the SDR ratio under the application of a magnetic field. Figure 4b presents the time-integrated PL intensity recorded under a  longitudinal external magnetic field (Faraday geometry) with a circularly  and linearly  polarized excitation. The triangles and the circles reproduce the  SDR ratios measured respectively with and without the external magnetic field. The SDR ratio grows significantly (from 140\% to 220\%) in the presence of the external magnetic field. It is interesting to note that 
the PL intensity under a linearly polarized excitation remains unchanged under the application of the external magnetic filed (not shown). The observed increase in the SDR
is entirely due to an enhancement of the PL intensity under circular excitation. This observation suggests that the external magnetic field leads to an improvement of the spin filtering mechanism.\\
Earlier~\cite{paget_1984} and more recent experiments~\cite{Kalevich_B_field_1,Kalevich_B_field_2} have investigated the increased efficiency of the SDR effect under a weak (100 mT) external magnetic field in dilute nitride GaAsN epilayers. The authors attributed the observed enhancement of the spin filtering effect to the suppression of the spin relaxation of the deep paramagnetic centers. This suppression was interpreted in terms of the spin orientation by hyperfine interaction of the nuclei in the deep centers leading to the dynamical nuclear polarization at room temperature.\\
The quantitatively and qualitatively very similar results obtained in this work
in implanted InGaAs are consistent with the nuclear  polarization of the Ga interstitial defects and support the assignment of the observed effects
to the spin dependent recombination.\\
In conclusion, we have demonstrated that active spin dependent recombination regions can be selectively created in N-free InGaAs epilayers by focus  ion beam implantation of Ga ions. We have evidenced a sizable SDR-driven effect reaching values as high as 240\% at low temperature. The application of a weak external magnetic field considerably enhances the spin filtering properties thanks to the spin polarization of nuclei surrounding the polarized paramagnetic centers. This put forward the use of ion implanted GaAs-based material as a model system for robust and long-lived spin memories in the form of nuclear spin polarization.
This demonstration, relying on the post-growth creation of Ga interstitial defects could be transferred to other Ga-based compounds allowing for the independent adjustment of PL wavelength and SDR efficiency.
The possibility of producing SDR active regions with arbitrary patterns by adjusting the implantation conditions could offer another degree of freedom for the design of possible spintronic devices relying on the spin filtering properties of deep paramagnetic defects.\\


\begin{thebibliography}{10}

\bibitem{lepine_1970}
D~L\'epine and J~J Pr\'ejean
\newblock Proc. 10th int. conf. phys. semicond.
\newblock In S~P Keller, J.~C. Hensel, and J.~Stern, editors, {\em Proceedings
  of the 10th International Conference on the Physics of Semiconductors}, page
  805, Boston, 1970

\bibitem{lepine_1972}
Daniel~J. Lepine
\newblock {\em Phys. Rev. B}, 6(2):436, 1972

\bibitem{wosinski_1977}
T~Wosinski, T~Figielski, and A~Makosa
\newblock {\em Phys. Status Solidi (a)}, 37(1):K57, 1976

\bibitem{solomon_1976}
I.~Solomon, D.~Biegelsen, and J.~C. Knights
\newblock {\em Solid State Commun.}, 22(8):505, 1977

\bibitem{weisbuch_1974}
C~Weisbuch and G~Lampel
\newblock {\em Solid State Commun.}, 14(2):141, 1974

\bibitem{paget_1984}
Daniel Paget
\newblock {\em Phys. Rev. B}, 30(2):931, 1984

\bibitem{kalevich:455}
VK~Kalevich, EL~Ivchenko, MM~Afanasiev, AY~Shiryaev, AY~Egorov, VM~Ustinov,
  B~Pal, and Y~Masumoto
\newblock {\em JETP Lett.}, 82(7):455, 2005

\bibitem{lombez:252115}
L.~Lombez, P.-F. Braun, H.~Carr\`{e}re, B.~Urbaszek, P.~Renucci, T.~Amand,
  X.~Marie, J.~C. Harmand, and V.~K. Kalevich
\newblock {\em Appl. Phys. Lett.}, 87(25):252115, 2005

\bibitem{kalevich:174}
V.~K. Kalevich, A.~Yu. Shiryaev, E.~L. Ivchenko, A.~Yu. Egorov, L.~Lombez,
  D.~Lagarde, X.~Marie, and T.~Amand
\newblock {\em JETP Lett.}, 85(3):174--178, 2007

\bibitem{lagarde:208}
D.~Lagarde, L.~Lombez, X.~Marie, A.~Balocchi, T.~Amand, V.~K. Kalevich,
  A.~Shiryaev, E.~Ivchenko, and A.~Egorov.
\newblock {\em Phys. Status Solidi (a)}, 204(1):208, 2007

\bibitem{zhao:241104}
F.~Zhao, A.~Balocchi, A.~Kunold, J.~Carrey, H.~Carrere, T.~Amand,
  N.~Ben~Abdallah, J.~C. Harmand, and X.~Marie
\newblock {\em Appl. Phys. Lett.}, 95(24):241104, 2009

\bibitem{Kalevich_B_field_1}
V.~K. Kalevich, M.~M. Afanasiev, A.~Yu. Shiryaev, and A.~Yu. Egorov
\newblock {\em Phys. Rev. B}, 85:035205, Jan 2012

\bibitem{Kalevich_B_field_2}
V.~K. Kalevich, M.~M. Afanasiev, A.~Y Shiryaev, and A.~Y. Egorov
\newblock {\em JETP Letters}, 96(9):567, 2013

\bibitem{wang_nature_materials}
X.~J. Wang, I.~A. Buyanova, F.~Zhao, D.~Lagarde, A.~Balocchi, X.~Marie, C.~W.
  Tu, J.~C. Harmand, and W.~M. Chen
\newblock {\em Nat. Mater.}, 8(3):198, 2009

\bibitem{weyers_red_1992}
Markus Weyers, Michio Sato, and Hiroaki Ando
\newblock {\em Japanese Journal of Applied Physics}, 31(Part 2, No.
  {7A}):L853, 1992

\bibitem{wu_band_2002}
J.~Wu, W.~Shan, and W.~Walukiewicz.
\newblock {\em Semiconductor Science and Technology}, 17(8):860, 2002

\bibitem{Zhao_JOP}
F.~Zhao, A.~Balocchi, G.~Truong, T.~Amand, X.~Marie, X.~J. Wang, I.~A.
  Buyanova, W.~M. Chen, and J.~C. Harmand.
\newblock {\em Journal of Physics: Condensed Matter}, 21(17):174211, 
  2009

\bibitem{kunold_giant_2011}
A.~Kunold, A.~Balocchi, F.~Zhao, T.~Amand, N.~Ben Abdallah, J.~C. Harmand, and
  X.~Marie
\newblock {\em Physical Review B}, 83(16):165202,  2011

\end{thebibliography}
\end{document}